# SPIN MICROSCOPE BASED ON OPTICALLY DETECTED MAGNETIC RESONANCE

Boris M. Chernobrod and Gennady P. Berman
Theoretical Division, Los Alamos National Laboratory, Los Alamos, NM 87545

## Abstract

We propose a scanning magnetic microscope which has a photoluminescence nanoprobe implanted in the tip of an AFM or STM, or NSOM, and exhibits optically detected magnetic resonance (ODMR). The proposed spin microscope has nanoscale lateral resolution and the single spin sensitivity for AFM and STM.

Continuing progress of nanotechnology including spintronics and quantum information processing, based on solid state quantum computer, has brought significant attention to the problem of measurements of magnetic properties of materials with sub-nanometer spatial resolution. The recently developed techniques that demonstrate the highest sensitivity and spatial resolution are magnetic resonance force microscopy (MRFM) and optically detected magnetic resonance (ODMR). Significant progress in MRFM has been made since the first experiment, which was performed at IBM by a team led by Rugar [1]. (See the detailed review [2].) Today MRFM promises to achieve single spin sensitivity with several nanometer spatial resolution. The main achievements in MRFM method are related to transfer of the detection of a very weak microwave signal to the detection of the mechanical oscillation of a micro-cantilever. Another option to enhance the sensitivity is transfer of the microwave signal to the optical domain, which is realized in the ODMR method. The related single spin experiments were independently performed in 1993 by two groups led by Moerner [3] and Orrit [4]. Today the principles of detection of a single spin based on ODMR are well established. (See review [5].) The limitation of the lateral resolution of ODMR is related to the size of the light spot. The highest resolution is obtained by a near-field scanning optical microscope (NSOM), which has a light spot size of about 30 – 50 nm. Another limitation of the ODMR technique is that the unpaired electron has to be a part of a molecule, which absorbs or emits light.

In the present paper we propose a modification of the ODMR technique which is free from the limitations of the conventional ODMR method. In our approach a photoluminescent nanoparticle or other photoluminescent center located in the tip apex exhibits ODMR in the vicinity of unpaired electron spins or nuclear magnetic moments in the sample. We propose several approaches to this spin microscope based on ODMR, the general layouts of which are shown in Fig. 1, 3-5. Fig. 1 presents a design based on the apertureless scanning optical microscope, which exploits the highly sensitive AFM tip modified by implanting a nano-size photoluminescent particle in the apex of the tip. The sample to be observed is located in close proximity to the tip-on-cantilever system, and a permanent magnet is placed nearby. A nearby radio-frequency coil produces an oscillating field at the frequency resonant with the transition between the magnetic sub-levels of the photoluminescent nanoparticle. Or another photoluminescent center located in the tip apex exhibits ODMR in the vicinity of unpaired electron spins or nuclear magnetic moments in the sample. The nano-probe absorption in an evanescent laser field could be significantly enhanced at the sharp apex of silicon tip. As it has been demonstrated by the JILA/NIST group [6], the combination of AFM with near-field optical scanning microscope (NSOM) method exhibits nano-scale spatial resolution and a significant increase in the detection sensitivity compared with the standard NSOM method. To avoid the restrictions related to the necessity to mount the sample on the prism, one uses a design in which the light comes through the fiber (Fig. 2). It could be a tapered fiber, which provides an evanescent field, or a regular fiber could be used, as well.

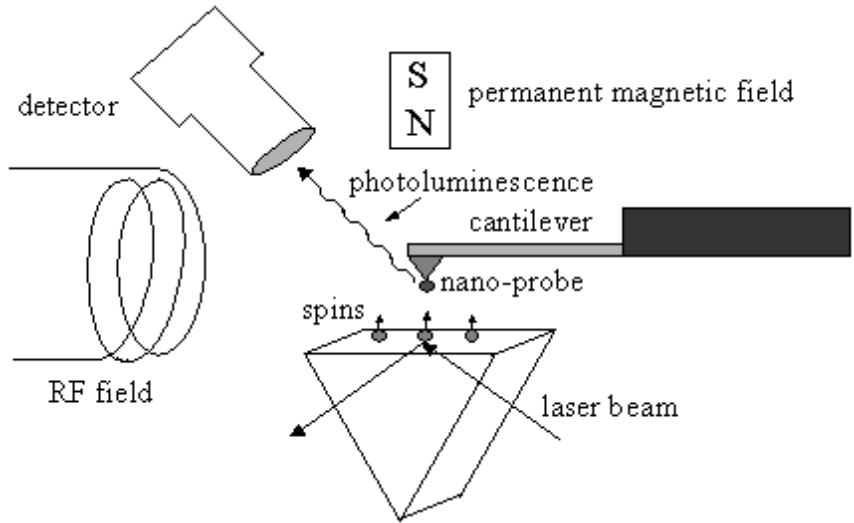

Figure 1: A schematic view of an ODMR-based scanning microscope combining the AFM and NSOM methods.

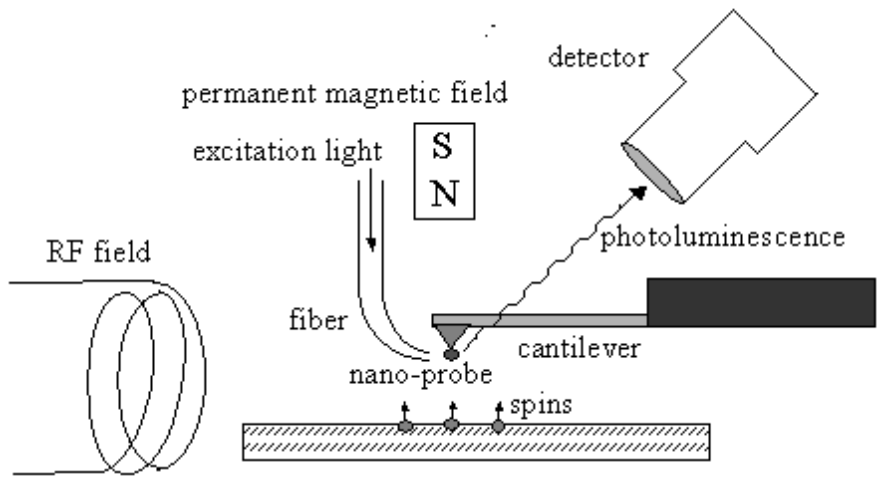

Figure 2: A schematic view of an ODMR-based scanning microscope illuminated through a fiber.

The resolution of this method is related to the size of the photoluminescent nano-probe, which is typically in the range of 1 – 10 nm, or even of Ångström-scale when a single fluorescent molecule is used as a probe. The application of a fluorescent probe for nanometer scale lateral resolution was first proposed and experimentally demonstrated by the team led by Letokhov [7,8]. In their approach, local fluorescent probes were employed for the fluorescent resonance energy transfer, where the donor molecules, located in the tip apex, were used to excite the fluorescence of an acceptor center of the sample. Recent progress in the photoluminescence efficiency of semiconductor nanoparticles has catalyzed a broad spectrum of applications of these particles as luminescent nano-probes. Semiconductor nanocrystals with nanometer diameters exhibit high quantum yields [9,10], typically over 50%, and high stability. In a semiconductor, quantum dot confinement leads to a replacement of continuous bands of energy by molecular-like energy levels structure [11]. As was demonstrated for single molecules and nanostructures [5,12] the sensitivity of the ODMR method to the external magnetic field is higher for narrower photoluminescence spectrum substructure.

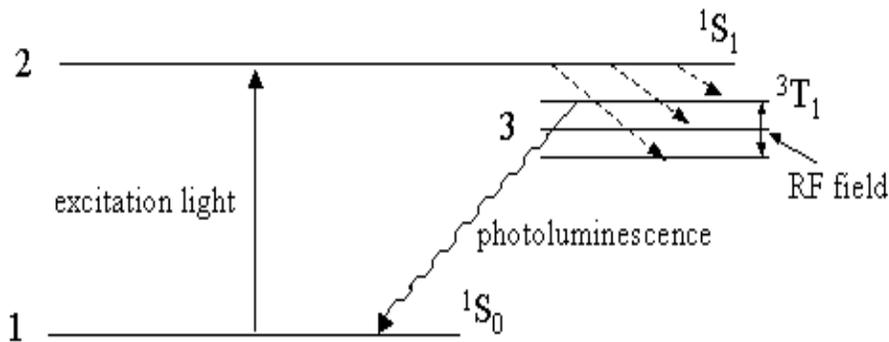

Figure 3: A schematic representation of the energy level diagram and ODMR-based transitions of a nanoparticle.

The theory of the spectrum of quantum dots [11] shows that the ODMR spectrum depends on the mutual hole-electron interaction in the exciton. The hole and electron spins create manifolds corresponding to singlet (S=0) or triplet (S = 1) spin state or other type manifolds depending on the anisotropy of interaction. The spectroscopic scheme of ODMR (Fig. 3) comprises the optical transition 1-2 corresponding to excitation with absorption of a photon from the laser field, the non-radiant transition to the manifold of the upper luminescent levels 3, and the luminescent transition 3 –1. The permanent magnetic field changes the energy splitting of magnetic sublevels, and the *rf* field induces transitions between sublevels, which are monitored as increasing or decreasing of the photoluminescence intensity. The sensitivity of our proposed scanning microscopy method is strongly dependent on the sharpness of the spectral structures measured in ODMR experiments. There are two possible types of measurements. In the first type of measurement the frequency of oscillating magnetic field is fixed and the external permanent magnetic field is varied. In the second type of measurement, the permanent magnetic field is fixed and the frequency of oscillating magnetic field is varied in the vicinity of the resonance frequency. Experiments with variable permanent magnetic field demonstrate rather sharp spectral structures in the range 0.1 – 0.002 T at sample temperatures of about 4º K [12,13]]. Another possibility is to use dye molecules as the nano-probe. It is well known that at low temperature dye molecules have a very narrow ODMR width of about $10^{-3}$ T [5]. Our preliminary analysis of the dependence of the sensitivity on geometry shows that at optimal conditions an ODMR of a nano-size probe can sense the magnetic field of a single electron spin. To estimate the sensitivity limit of an ODMR-based scanning microscope, we assume that the diameter of the nano-probe is 1 nm; the distance between the nano-probe edge and the surface is 5 Å; and in a radius of several nm there is only one unpaired spin. This spin is oriented perpendicular to the surface. The magnetic field from the single spin is given by $B_s = (\mu_0/4\pi)(3\mathbf{n}(\mathbf{mn}) - \mathbf{m})/r^3$. When we substitute $\mathbf{m} = -(1/2)g_e\mu_b = -9.28 \times 10^{-24}$ J/T, $\mu_0/4\pi = 10^{-7}$ N/A$^2$, we get $B_s = 1.5 \times 10^{-2}$ T. This value of the magnetic field of individual spin is larger than the most narrow features of the ODMR spectrum, and could be detected by measurement of the resonance shift. Note that the procedure of measurement of the static magnetic field of a single spin does not perturb its quantum state and could be considered as an example of a non-demolition measurement of a quantum object. This property is very desirable for quantum computing applications, where the two states of an individual spin represent a qubit.

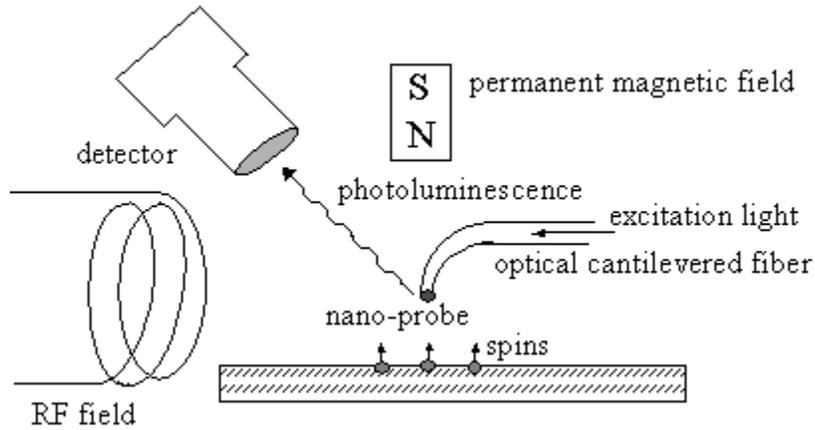

Figure 4: A schematic view of the ODMR-based scanning microscope based on an optical cantilevered fiber.

The scheme with spatial resolution of several tens of nanometers is presented in Fig. 4. The cantilevered optical fiber tip developed by NANONICS Imaging is modified by implantation of a photoluminescent nanoparticle in the apex of tip. This scheme could be used in combination with an opaque sample. The resolution of this scheme is limited by the accuracy of positioning of the cantilevered fiber, which could be in a range of tens of nanometers.

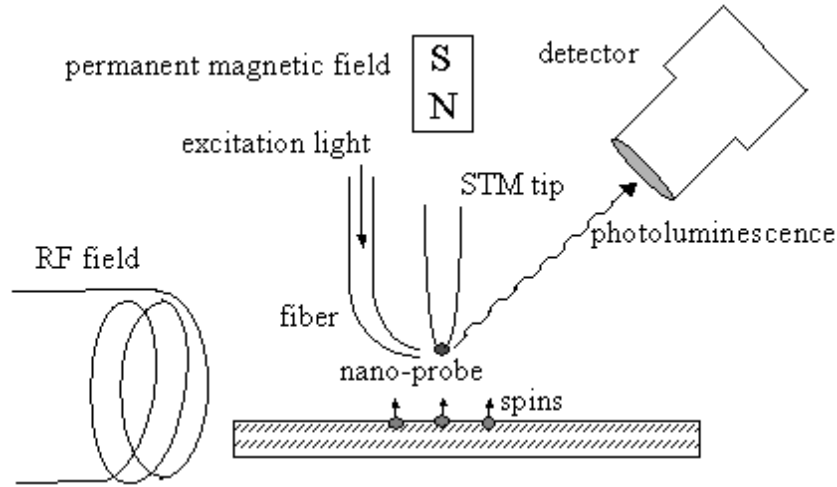

Figure 5: A schematic view of ODMR-based scanning microscope based on the STM method.

Fig. 5 presents a scheme which uses the STM. The STM-tip is has a nano-probe attached. The sample is characterized by its conductivity. The resolution of this method is limited by the size of the nano-probe.

In conclusion we propose a scanning magnetic microscope which incorporates a photoluminescence nanoprobe implanted in the apex of the tip of the AFM or the STM, or the NSOM microscope, and exhibits ODMR. The proposed technique has nanoscale lateral resolution and single spin sensitivity in the cases of modified AFMs and STMs.

This work was supported by the Department of Energy (DOE) under Contract No. W-7405-ENG-36, by the National Security Agency (NSA), and by the Advanced Research and Development Activity (ARDA).